\documentclass[aps, prb, reprint, nofootinbib, superscriptaddress, noshowpacs,floatfix]{revtex4-2}
\usepackage{microtype}
\usepackage[T1]{fontenc}
\usepackage[utf8]{inputenc}
\usepackage[english]{babel}
\usepackage{amssymb}
\usepackage{amsmath}
\usepackage{graphicx}
\usepackage{bm}
\usepackage[hidelinks]{hyperref}

\usepackage{siunitx}
\DeclareSIUnit\meter{\text{m}}
\DeclareSIUnit\second{\text{s}}
\DeclareSIUnit\ampere{\text{A}}
\DeclareSIUnit\kelvin{\text{K}}
\DeclareSIUnit\mole{\text{mol}}
\DeclareSIUnit\coulomb{\text{C}}
\DeclareSIUnit\farad{\text{F}}
\DeclareSIUnit\henry{\text{H}}
\DeclareSIUnit\joule{\text{J}}
\DeclareSIUnit\ohm{\text{$\Omega$}}
\DeclareSIUnit\volt{\text{V}}
\DeclareSIUnit\watt{\text{W}}
\DeclareSIPrefix\deca{\text{da}}{1}
\DeclareSIPrefix\hecto{\text{h}}{2}
\DeclareSIPrefix\kilo{\text{k}}{3}
\DeclareSIPrefix\mega{\text{M}}{6}
\DeclareSIPrefix\giga{\text{G}}{9}
\DeclareSIPrefix\tera{\text{T}}{12}
\DeclareSIPrefix\peta{\text{P}}{15}
\DeclareSIPrefix\exa{\text{E}}{18}
\DeclareSIPrefix\zetta{\text{Z}}{21}
\DeclareSIPrefix\yotta{\text{Y}}{24}
\DeclareSIPrefix\deci{\text{d}}{-1}
\DeclareSIPrefix\centi{\text{c}}{-2}
\DeclareSIPrefix\milli{\text{m}}{-3}
\DeclareSIPrefix\micro{\text{$\mu$}}{-6}
\DeclareSIPrefix\nano{\text{n}}{-9}
\DeclareSIPrefix\pico{\text{p}}{-12}
\DeclareSIPrefix\femto{\text{f}}{-15}
\DeclareSIPrefix\atto{\text{a}}{-18}
\DeclareSIPrefix\zepto{\text{z}}{-21}
\DeclareSIPrefix\yocto{\text{y}}{-24}
\allowdisplaybreaks

\graphicspath{ {./figures/} }

\renewcommand\vec{\mathbf}

\newcommand{\ud}{\,{\mathrm{d}}}
\newcommand{\unabla}{\vec{\nabla}}

\newcommand{\uL}{L}  

\newcommand{\uX}{X}  
\newcommand{\uex}{\vec{e}_\mathrm{X}}  
\newcommand{\ux}{x}  

\newcommand{\ut}{t}  

\newcommand{\uC}{C}  
\newcommand{\uCmax}{C_\mathrm{max}} 
\newcommand{\uc}{c} 
\newcommand{\ucx}{c_{x}} 
\newcommand{\ucZ}{b} 
\newcommand{\ucZx}{b_{x}} 
\newcommand{\ur}{r}

\newcommand{\up}{p} 
\newcommand{\ualpha}{\alpha} 

\newcommand{\uD}{D} 
\newcommand{\uT}{T} 
\newcommand{\ukB}{k_\mathrm{B}} 
\newcommand{\ukBT}{\ukB\uT} 

\newcommand{\uvJ}{\vec{J}} 
\newcommand{\uje}{{\cal I}} 

\newcommand{\uq}{q} 
\newcommand{\urho}{\rho} 

\newcommand{\uvE}{\vec{E}} 

\newcommand{\uPhi}{\varPhi} 
\newcommand{\uphi}{\varphi} 
\newcommand{\uphix}{\varphi_{x}} 
\newcommand{\uphixx}{\varphi_{xx}} 

\newcommand{\ueps}{\epsilon} 
\newcommand{\uepsZ}{\epsilon_{0}} 
\newcommand{\uepsepsZ}{\ueps\uepsZ} 

\newcommand{\ul}{\varLambda} 
\newcommand{\ulZ}{\varLambda_0} 
\newcommand{\ulxx}{\varLambda_{xx}} 

\newcommand{\uAconst}{{\cal A}} 

\newcommand{\uupsilon}{\upsilon}

\begin{document}
\selectlanguage{english}
\title{Exact solution for stationary states of a closed memristor with mobile charged vacancies}
\author{\firstname{I.~V.} \surname{Boylo}}
\email[]{boylo@donfti.ru}
\affiliation{A.A. Galkin Donetsk Institute for Physics and Engineering, Roza Luxembourg str. 72, Donetsk, Russia 283048}
\author{\firstname{K.~L.} \surname{Metlov}}
\email[]{metlov@donfti.ru}
\affiliation{A.A. Galkin Donetsk Institute for Physics and Engineering, Roza Luxembourg str. 72, Donetsk, Russia 283048}
\date{\today}
\begin{abstract}
A nonlinear model of a memristor based on charged mobile vacancies is considered taking into account the electrostatic interaction between them. This interaction significantly affects the stationary (limiting) vacancy distributions formed under the action of the electric current flowing through the memristor, for which analytical expressions are obtained in this work. Between the regions with reduced and increased vacancy concentrations, an intermediate electrically neutral region is formed due to the electrostatic interaction. Interestingly, the limiting resistances in the ``on'' and ``off'' states of such a memristor do not depend on the strength of the electrostatic interaction, at least in the leading first order in the vacancy concentration.
\end{abstract}
\maketitle
\section{Introduction}
The concept of the memristor was proposed by Leon Chua~\cite{Chua1971,Chua1976} in the 1970s as a logical completion of the set of passive two-terminal elements of electrical circuits. Unlike a resistor, whose voltage drop is determined by the instantaneous value of the electric current, the resistance of a memristor is determined by the time integral of the current---the total electric charge that has passed through the memristor---and thus stores an imprint of the history of current changes in the past. In other words, a memristor is a resistor with memory.

The first physical realization of a memristor was created in 2008 at Hewlett-Packard~\cite{strukov2008} and used the electric-current-controlled motion of oxygen vacancies in TiO$_2$. The state of the memristor changes due to the redistribution of vacancies, which are driven by the electron current. Strictly speaking, the term ``memristor'' implies that the resistance depends only on the charge that has passed. There also exists a broader class of memristive systems~\cite{Chua1976} that also exhibit hysteresis of current--voltage characteristics but whose resistance may depend explicitly on time. These include systems with resistance switching due to the growth of conducting filaments that close the electrical circuit in NiO~\cite{bruyere1970}, other memristive systems based on transition-metal oxides~\cite{Sawa2008}, heterojunctions based on perovskite compounds~\cite{fujii2004}, and nanoscale electrolytic heterostructures~\cite{emmerich2024}.

Initially, memristive devices were intended as a realization of nonvolatile random-access memory, but today flash memory leads in this competition. However, the recent development of artificial-intelligence systems has revealed the shortcomings of the von Neumann computer architecture, in which the processor and memory are separate functional units connected by a data bus. The implementation of neural networks, first, requires large amounts of memory, and, second, practically all of this memory is accessed at every iteration of the network operation. Within the von Neumann architecture this requires a huge memory bus bandwidth, while the random access speed to an individual memory cell is of little importance. Nature has solved an analogous problem differently by combining the functions of information transmission and storage in the synapses connecting neurons. Modern memristive devices can play the same role; see, for example, the review~\cite{shooshtari2025}.

Modeling of memristor operation is mainly performed numerically~\cite{Strukov2009,Rozenberg2010,rozen2010,Larentis2012,Kim2014,Marchewka2016,Boylo2017,agudov2020,agudov2021} within rather complex multi-parameter models. Such models reproduce experimental data well, but a comprehensive study of the memristor problem, making it possible to distinguish individual physical effects and investigate limiting cases, is possible only within an analytical approach. For this purpose, we have proposed a nonlinear memristor model~\cite{boylo2020} based on mobile vacancies that admits (in its original formulation, which reduces to the Burgers equation) a complete analytical solution. It describes the state of the memristor (determined by the vacancy distribution in it) much more accurately than the widely used two-phase model~\cite{strukov2008}. In the present work we generalize the model of Ref.~\cite{boylo2020} by including the non-local electrostatic interaction between charged vacancies.

\section{Model}

\begin{figure}
\begin{center}
\includegraphics[width=1.0\columnwidth]{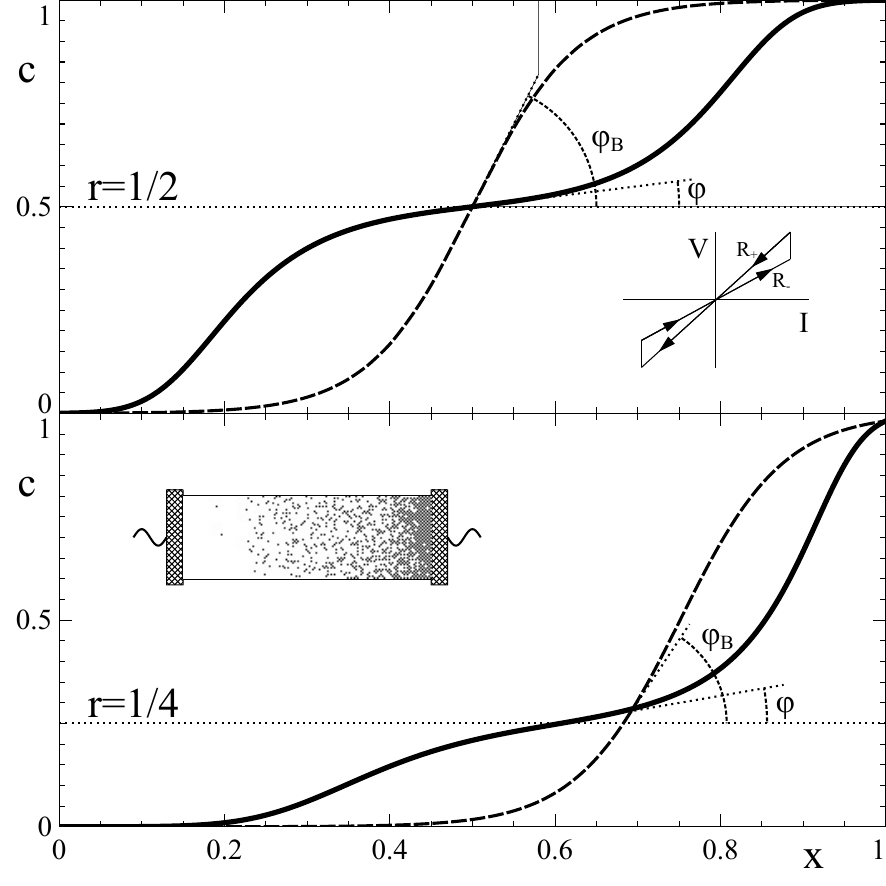}
\vspace{-0.8cm}
\end{center}
\caption{\label{fig:profiles} Stationary vacancy distributions in memristors half-filled with vacancies ($r=1/2$, top) and quarter-filled ($r=1/4$, bottom) at normalized current strength $p=16$; solid lines show distributions at normalized Coulomb energy of vacancies $\alpha=1000$, dashed lines at $\alpha=0$, corresponding to solutions of the Burgers equation. The insets show: (top right) a typical hysteresis loop of the considered memristor under a square-wave current; (bottom left) an illustration of the random vacancy distribution in the memristor corresponding to the solid line in the figure. The derivative scaling factor is defined in the text as $\uupsilon=\tan\varphi/\tan\varphi_\mathrm{B}$.}
\end{figure}

Consider a one-dimensional memristor of length $\uL$ containing mobile electrically charged vacancies, whose local concentration $\uC$ depends on one spatial coordinate $0<\uX<\uL$. Let the memristor be completely closed and not exchanging atoms with the environment (it exchanges only electrons). Then the motion of oxygen vacancies in it obeys the continuity equation: $\partial\uC/\partial\ut=-\unabla\cdot\uvJ$, where for the vacancy current we will use the model:
\begin{equation}
 \label{eq:vacCurrent}
 \uvJ = -\uD\unabla\uC + \uC \left(1-\frac{\uC}{\uCmax}\right)\frac{2\uD\uq\uvE}{\ukBT},
\end{equation}
which follows from the model of Ref.~\cite{boylo2020} for the thermal drift of vacancies in a memristor under the action of the electric field $\uvE$ in the continuous-drift regime ($a\rightarrow0$ in the notation of Ref.~\cite{boylo2020}). Here $\uD$ is the diffusion coefficient [\si{\square\meter\per\second}], $\uCmax$ is the maximum achievable concentration of vacancies, measured (like $\uC$) in [\si{\mole\per\cubic\meter}], $\uq$ is the electric charge of one vacancy in [\si{\coulomb}], $\uvE$ is the local electric field in [\si{\volt\per\meter}], $\ukB$ is the Boltzmann constant [\si{\joule\per\kelvin}], and $\uT$ is the absolute temperature [\si{\kelvin}]. The electric field $\uvE = \urho \uje \uex - \unabla\uPhi$ consists of two terms: the first is the constant field caused by the direct electric current with density $\uje$ [\si{\ampere\per\square\meter}] flowing through the memristor with material resistivity $\urho$ [\si{\ohm\meter}]; and the second is the electric field created by the electric charges of the vacancies themselves. The potential $\uPhi$ is determined by the Poisson equation
\begin{equation}
\label{eq:Poisson}
\Delta\uPhi = - \frac{\uq}{\uepsepsZ} (\uC - \uC_0),
\end{equation}
where $\uC_0$ is the vacancy concentration corresponding to the overall electrical neutrality of the whole memristor, and $\uepsepsZ$ is the permittivity of the memristor material [\si{\farad\per\meter}]. Since well-conducting contacts are connected to the memristor (as shown schematically in the inset to the lower part of Fig.~\ref{fig:profiles}), the potential $\uPhi|_{\uX=0,\uL}=0$. Until now the potential $\uPhi$ was not taken into account in the considered non-linear memristor model, its inclusion is the main feature of the present work.

The total electrical resistance of the memristor in the model of Ref.~\cite{boylo2020} is calculated phenomenologically, taking into account the linear (first order in Taylor's expansion) term in the dependence of the resistivity (bulk and surface at the contacts) on the local vacancy concentration. In this order, the redistribution of vacancies in a closed memristor (with their total number conserved) does not change the total resistance of its bulk (although it does affect the resistance switching kinetics~\cite{boylo2025_en}), and only the surface term contributes to the change of the total resistance: $R=R_0+(R_1+R_2)\sigma$, where~\cite{boylo2025_en}
\begin{equation}
 \label{eq:resistance}
 \sigma=\frac{\uC|_{\uX=0}}{\uCmax} \cos^2\theta + \frac{\uC|_{\uX=L}}{\uCmax} \sin^2\theta,
\end{equation}
where $R_0$, $R_1$, $R_2$ are parameters of the memristor material and contacts [\si{\ohm}], and the angle $0<\theta=\arctan \sqrt{R_2/R_1} < \pi/2$ characterizes the asymmetry of the interfaces. For the resistance of a closed memristor in the ``on'' and ``off'' states to be different, its interfaces (contact materials and/or structure) must be different.

\section{Stationary states of the memristor}
Unfortunately, it is not yet possible to solve exactly the full kinetic equation following from Eqs.~\eqref{eq:vacCurrent} and~\eqref{eq:Poisson}. This equation is similar to the Burgers equation but contains a nonlocal integral term expressing the derivative of the electrostatic potential $\uphi$ through the Green's function of the Poisson equation. It cannot be linearized by the Cole--Hopf substitution.

However, the stationary states of the memristor are also of considerable interest, since it is precisely these vacancy distributions that determine the resistances in the ``on'' and ``off'' states. In a closed memristor these distributions correspond to the complete absence of the vacancy current $\uvJ=0$, which follows from the continuity equation with the boundary conditions taken into account. In the one-dimensional case under consideration this reduces to the following dimensionless system of equations
\begin{subequations}
\label{eq:stationary}
\begin{align}
 &-\ucx + \uc(1-\uc) \left(\up - \left(\alpha/2\right)\uphix\right) = 0, \label{eq:jzero}\\
 &\uphixx = - (c - r), \label{eq:phiPoisson}
\end{align}
\end{subequations}
where $\uc=\uC/\uCmax$ and $\uphi=\uPhi \uepsepsZ/(\uq\uCmax\uL^2)$ are functions of the dimensionless coordinate $\ux=\uX/\uL$; the dimensionless parameter $\up=2\uq\urho\uje\uL/(\ukBT)$ characterizes the strength of the driving current density $\uje$ flowing through the memristor, and the dimensionless parameter $\ualpha = 4\uq^2\uL^2\uCmax/(\uepsepsZ\ukBT)>0$ characterizes the strength of the interaction between the vacancy charges. The filling factor
\begin{equation}
 \label{eq:r}
 \ur=\int_0^1 c(x)\ud x
\end{equation}
is an integral of motion in a closed memristor. For $\ualpha=0$ the problem reduces to finding the stationary solution of the Burgers equation, which is well known:
\begin{equation}
\uc|_{\ualpha=0}=\ucZ=\frac{e^{\up\ux}(e^{\up\ur}-1)}{e^{\up}-e^{\up\ur}-e^{\up\ux}+e^{\up(\ur+\ux)}}.
\end{equation}

To find the solution for $\ualpha\neq0$, we introduce a new function (the logit)
\begin{equation}
 \ul = \log\frac{\uc}{1-\uc},
\end{equation}
then the flux equation~\eqref{eq:jzero} is linearized and easily integrated ($\ul=p x - \ualpha\uphi/2 +\ulZ$), while Eq.~\eqref{eq:phiPoisson} becomes
\begin{equation}
\ulxx = \frac{\ualpha}{2}\left(
\frac{e^\ul}{1+e^\ul} - \ur
\right).
\end{equation}
Although this equation is nonlinear, it is autonomous, and its analytical solution can be expressed as a quadrature:
\begin{equation}
\label{eq:sol}
\int_{\ulZ}^\ul \frac{1}{\uAconst + \ualpha (\log (1+e^t) - \ur t )} \ud t = x,
\end{equation}
where two integration constants enter: $\ulZ=\ul|_{x=0}$ and $\uAconst$. It is convenient to define the latter in the form
\begin{equation}
 \uAconst = p^2\uupsilon^2 + \ualpha
 \left[\left(1-\ur\right)\log\left(1-\ur\right) + \ur \log \ur\right]
\end{equation}
through another constant $\uupsilon$, which has a simple physical meaning and is defined as
\begin{equation}
 \uupsilon = \frac{\ucx|_{\uc=\ur}}{\ucZx|_{\ucZ=\ur}}
\end{equation}
the derivative of the vacancy distribution profile at the point $\uc=\ur$, normalized to the analogous derivative in the absence of electrostatic interaction ($\ualpha=0$). Generally speaking, the coordinates $\ux$ of these two points with $\uc=r$ are different, as is clearly seen in Fig.~\ref{fig:profiles}. It can be shown that always $0<\uupsilon<1$, which makes this parameter especially convenient. The obtained general solution depends on two integration constants $\ulZ$ and $\uupsilon$. The boundary condition for the potential $\uphi(0)=0$ is then satisfied automatically, and from the condition $\uphi(1)=0$ it follows that $\ul(1)-\ul(0)=p$. The second equation for the integration constants follows from the conservation of the total number of vacancies~\eqref{eq:r}. The solution for $\ulZ$ can be obtained explicitly:
\begin{equation}
 \label{eq:ulZ}
 \ulZ = \log\frac{e^{\up\ur}-1}{e^\up-e^{\up\ur}},
\end{equation}
while the equation for $\uupsilon$, which is follows from Eq.~\eqref{eq:sol} by substituting $\ux=1$ and Eq.~\eqref{eq:ulZ}, can be solved only numerically. Examples of the resulting vacancy distribution profiles are shown in Fig.~\ref{fig:profiles}; they satisfy both Eqs.~\eqref{eq:stationary} and all boundary conditions for a given total number of vacancies in the memristor. One can also obtain an approximate (but sufficiently accurate for estimates) explicit expression for $\uupsilon$, given in Appendix~\ref{sec:upsapprox}.

\begin{figure}
\begin{center}
\includegraphics[width=1.0\columnwidth]{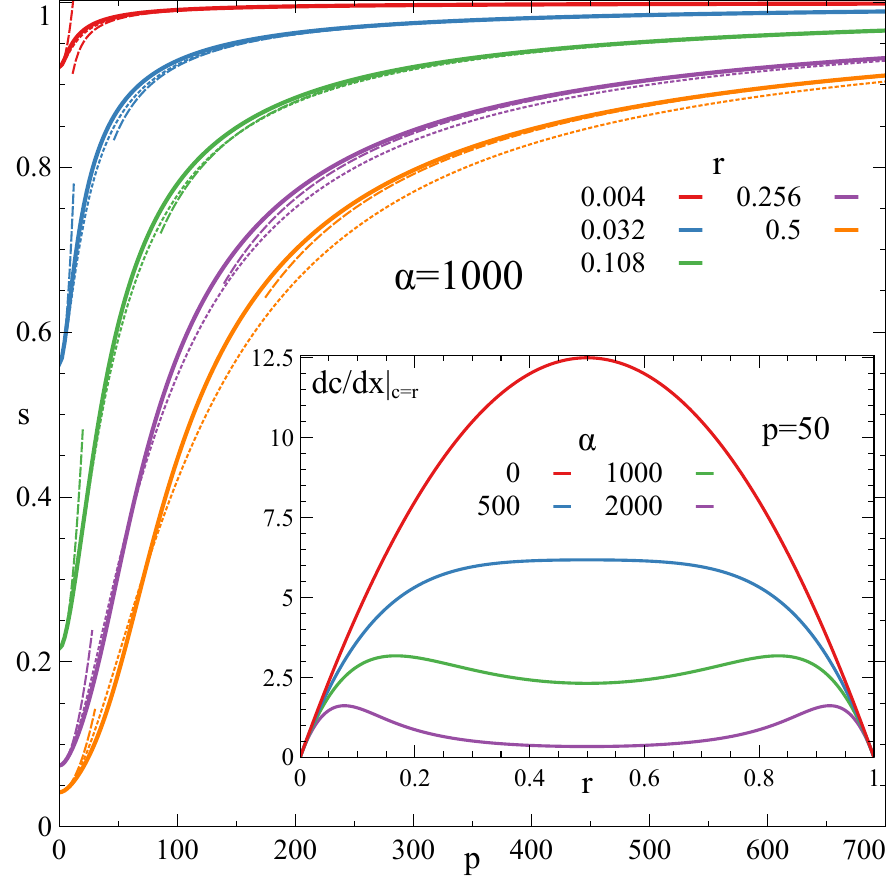}
\vspace{-0.8cm}
\end{center}
\caption{\label{fig:scale} Derivative scaling factor $s$ as a function of the reduced current strength $p$ in memristors with different vacancy fillings at $\alpha=1000$. The curves for fillings $r$ and $1-r$ coincide. The inset shows the dependence of the derivative of the vacancy distribution profile $\ud c/\ud x$ at the point $c=r$ as a function of the memristor filling factor $r$ at $p=50$.}
\end{figure}

The dependence of the derivative scaling factor $\uupsilon$ on the current strength $\up$ and on the degree of filling of the memristor with vacancies for some values of the electrostatic interaction strength $\ualpha$ is shown in Fig.~\ref{fig:scale}.
These dependences are symmetric with respect to half filling (fillings $\ur$ and $1-\ur$ lead to the same values of $\uupsilon$). By definition, the pure Burgers memristor with $\ualpha=0$ corresponds to $\uupsilon=1$. The same limiting value is approached by $\uupsilon$ as the current strength $\up$ increases, whose effect eventually exceeds the effect of the electrostatic repulsion.

In general, the presence of electrostatic interaction leads to the fact that in the stationary state of the memristor, instead of two regions---one enriched and one depleted in vacancies---three regions are formed: an enriched region, a depleted region, and an intermediate electrically neutral region (the more extended the higher $\ualpha$). It seems (although the authors did not emphasize this) that such a transition region is observed in memristors based on oxygen vacancies in argon-ion-implanted rutile~\cite{SVNK2023}.

However, despite the fact that the vacancy distribution changes strongly under the influence of the electrostatic interaction between them, this interaction does not affect the vacancy concentrations at the ends of the memristor ($\uc(0)$ and $\uc(1)$), which follows from Eq.~\eqref{eq:ulZ} and the independence of $\ulZ$ from $\ualpha$. This means that the limiting resistances (in the ``on'' and ``off'' states) of such a memristor in the linear approximation~\eqref{eq:resistance} also do not depend on the strength of the electrostatic interaction.

\section{Conclusions}
A nonlinear memristor model based on mobile vacancies in a material where the vacancy concentration assumes values in a finite fixed range is extended by taking into account the electrostatic interaction between the vacancies. Analytical expressions are obtained for the stationary (at a certain value of the current flowing through the memristor) vacancy distributions in the memristor. It is shown that the electrostatic interaction does not affect the electrical resistance of the memristor in the ``on'' and ``off'' states, at least in the leading linear order in the vacancy concentration in the bulk and at the contacts.

\appendix
\renewcommand{\appendixname}{Appendix}
\renewcommand{\thesection}{\Alph{section}}
\section{Approximate expression for $\uupsilon$}
\label{sec:upsapprox}

For $\uupsilon$ the following asymptotic expressions can be obtained:
\begin{align}
 \label{eq:upssmall}
 \upsilon (p\ll1) = & \uupsilon(0) + \frac{\uupsilon''(0)}{2!} p^2 + O(p^3), \\
 \upsilon(0) = & \frac{h}{\sinh h}, \\
 \upsilon''(0) = & \frac{h \left(8 h^2-\ualpha \right) (6 h \coth h-\cosh 2 h -5)}{288 \ualpha \sinh^3 h}, \\
 \label{eq:upslarge}
 \upsilon (p\gg1) = & 1 - \frac{2h^2}{p} + O(p^{-2}),
\end{align}
where $h=\sqrt{\ualpha \ur (1-\ur)}/(2\sqrt{2})$.

On their basis one can construct a rational approximation
\begin{align}
 \label{eq:upspade}
 \uupsilon(p) \approx & \frac{h(576+a \up (1 + (\ur-1) \ur)))}{\sinh h (576+b \up (1 + (\ur-1) \ur)))}, \\
 a/b =& \frac{\sinh h \left(2 h^3-h \up+\up \sinh h\right)}{h \left(\left(2 h^2+\up\right) \sinh h-h \up\right)}, \\
 a-b =& 2 p-\frac{6 \up (h \coth h-1)}{\sinh^2 h},
\end{align}
which, although not exactly, nevertheless approximates the values of $\uupsilon$ sufficiently well over the whole parameter range of the problem. The asymptotics~\eqref{eq:upssmall}, \eqref{eq:upslarge}, and the approximate expression~\eqref{eq:upspade} are shown in Fig.~\ref{fig:scale} by dashed and dotted lines.

%
\end{document}


\selectlanguage{russian}
\title{Точное решение для стационарных состояний закрытого мемристора с подвижными заряженными вакансиями}
\author{\firstname{И.~В.} \surname{Бойло}}
\email[]{boylo@donfti.ru}
\affiliation{Донецкий Физико-Технический Институт им. А.А. Галкина, г. Донецк, ДНР, Россия 283048}
\author{\firstname{К.~Л.} \surname{Метлов}}
\email[]{metlov@donfti.ru}
\affiliation{Донецкий Физико-Технический Институт им. А.А. Галкина, г. Донецк, ДНР, Россия 283048}
\date{\today}
\begin{abstract}
Рассмотрена нелинейная модель мемристора на заряженных подвижных вакансиях с учётом электростатического взаимодействия между ними. Такое взаимодействие существенно влияет на стационарные (предельные) распределения вакансий, формирующиеся под действием проходящего через мемристор электрического тока, для которых в работе получены аналитические выражения. Между областями с пониженной и повышенной концентрациями вакансий, под действием элестростатического взаимодействия формируется промежуточная электрически нейтральная область. Интересно, что предельные сопротивления во ``включенном'' и ``выключенном'' состояниях такого мемристора от силы электростатического взаимодействия не зависят, по крайней мере в лидирующем первом порядке по концентрации вакансий.
\end{abstract}
\maketitle
\section{Введение}
Концепция мемристора была предложена Леоном Чуа~\cite{Chua1971,Chua1976} в 1970-х, как логическое завершение набора пассивных двухполюсных элементов электрических цепей. В отличие от резистора, падение напряжения на котором определено мгновенной величиной электрического тока, сопротивление мемристора определяется интегралом от тока по времени --- прошедшим через мемристор полным электрическим зарядом, т.е. хранит отпечаток истории изменения тока в прошлом. Другими словами, мемристор --- резистор с памятью.

Первая физическая реализация мемристора была создана в 2008-м на фирме Hewlett Packard~\cite{strukov2008} и использовала управляемое электрическим током перемещение кислородных вакансий в TiO$_2$. Состояние мемристора при этом изменяется за счёт перетока вакансий, а считывание и формирование этого состояния осуществляется посредством тока электронов. В строгом смысле, термин ``мемристор'' подразумевает, что сопротивление зависит только от прошедшего заряда. Существует и более широкий класс мемристивных систем~\cite{Chua1976} также демонстрирующих гистерезис вольт-амперных характеристик, но сопротивление которых может явно зависеть от времени. К таким относятся системы с переключением сопротивления за счёт роста замыкающих электрическую цепь проводящих нитей в NiO~\cite{bruyere1970}, другие мемристивные системы на основе оксидов переходных металлов~\cite{Sawa2008}, гетеропереходы на основе перовскитных соединений~\cite{fujii2004}, наномасштабные электролитические гетероструктуры~\cite{emmerich2024}.

Изначально мемристивные устройства планировались как реализация энергонезависимой памяти с произвольным доступом, но в этой конкурентной борьбе на сегодня лидирует флеш-память. Однако, развитие систем с искусственным интеллектом в последнее время проявляет недостатки фон-Неймановской архитектуры построения ЭВМ, в которой процессор и память являются отдельными функциональными блоками, связанными шиной передачи данных. Реализация нейросетей, во-первых, требует больших объемов памяти, а во-вторых, практически вся эта память опрашивается на каждой итерации работы нейросети. В рамках фон-Неймановской архитектуры это требует огромной пропускной способности шины памяти, скорость же произвольного доступа к отдельной ячейке памяти при этом большого значения не имеет. Природа решила аналогичную задачу по-другому, объединив функции передачи и хранения информации в соединяющих нейроны синапсах. Такую же функцию могут играть и современные мемристивные устройства, см. например обзор~\cite{shooshtari2025}.

В основном, моделирование работы мемристоров выполняется численно~\cite{Strukov2009,Rozenberg2010,rozen2010,Larentis2012,Kim2014,Marchewka2016,Boylo2017,agudov2020,agudov2021} в рамках довольно сложных многопараметрических моделей. Такие модели хорошо воспроизводят экспериментальные данные, но полное исследование задачи о мемристоре, позволяющее выделить отдельные физические эффекты и исследовать предельные случаи, возможно лишь в рамках аналитического подхода. Для этого нами была предложена нелинейная модель мемристора~\cite{boylo2020} на подвижных вакансиях, допускающая (в её исходной формулировке, сводящейся к уравнению Бюргерса) полное аналитическое решение. Она гораздо точнее описывает состояние мемристора (определяемое распределением вакансий в нём), по сравнению с широко используемой двухфазной моделью~\cite{strukov2008}. В данной работе мы обобщаем модель~\cite{boylo2020} путём дополнительного точного учёта нелокального электростатического взаимодействия между заряженными вакансиями.

\section{Модель}

\begin{figure}
\begin{center}
\includegraphics[width=1.0\columnwidth]{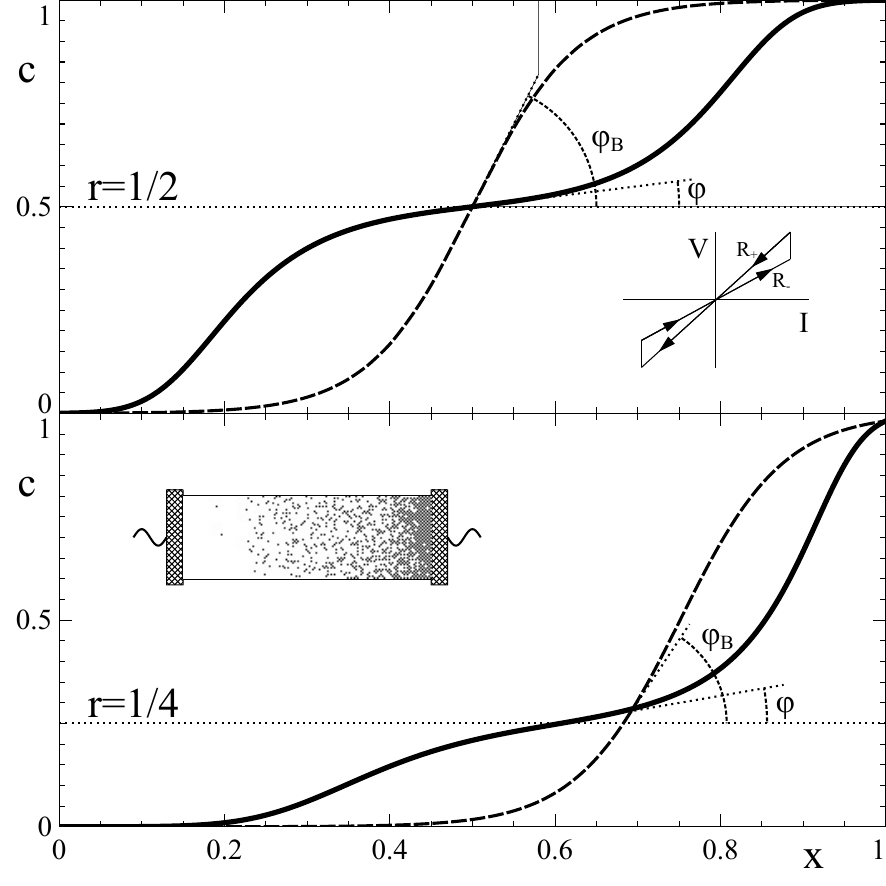}
\vspace{-0.8cm}
\end{center}
\caption{\label{fig:profiles} Стационарные распределения вакансий в мемристорах, заполненных вакансиями наполовину (r=1/2, вверху) и на четверть(r=1/4, внизу) при нормированной силе тока $p=16$; сплошными линиями показаны распределения при нормированной кулоновской энергии вакансий $\alpha=1000$, пунктирными при $\alpha=0$, соответствующие решениям уравнения Бюргерса. На вставках показаны: (справа вверху) типичная петля гистерезиса рассматриваемого мемристора при подаче на него меандра тока; (слева внизу) иллюстрация случайного распределения вакансий в мемристоре, соответствующего сплошной линии на рисунке. Коэффициент масштабирования производной определён в тексте как $\uupsilon=\tan\varphi/\tan\varphi_\mathrm{B}$.}
\end{figure}

Рассмотрим одномерный мемристор длины $\uL$, содержащий подвижные электрически заряженные вакансии, локальная концентрация которых $\uC$ зависит от одной пространственной координаты $0<\uX<\uL$. Пусть мемристор полностью закрыт и не обменивается атомами с окружающей средой (он обменивается только электронами). Тогда движение кислородных вакансий в нём подчиняется уравнению непрерывности: $\partial\uC/\partial\ut=-\unabla\cdot\uvJ$, где для тока  вакансий мы будем использовать модель:
\begin{equation}
 \label{eq:vacCurrent}
 \uvJ = -\uD\unabla\uC + \uC \left(1-\frac{\uC}{\uCmax}\right)\frac{2\uD\uq\uvE}{\ukBT},
\end{equation}
которая следует из модели~\cite{boylo2020} для теплового дрейфа вакансий в мемристоре под действием электрического поля $\uvE$ в режиме непрерывного дрейфа ($a\rightarrow0$ в обозначениях~\cite{boylo2020}). Здесь $\uD$ --- коэффициент диффузии [\si{\square\meter\per\second}], $\uCmax$ --- предельно достижимая (максимальная) концентрация вакансий, измеряемая (как и $\uC$) в [\si{\mole\per\cubic\meter}], $\uq$ --- электрический заряд одной вакансии в [\si{\coulomb}], $\uvE$ --- локальное электрическое поле в [\si{\volt\per\meter}], $\ukB$ --- постоянная Больцмана [\si{\joule\per\kelvin}] и $\uT$ --- абсолютная температура [\si{\kelvin}]. Электрическое поле $\uvE = \urho \uje \uex - \unabla\uPhi$ состоит из двух слагаемых: первое --- постоянное поле, вызванное протекающим через мемристор с удельным сопротивлением материала $\urho$ [\si{\ohm\meter}] постоянным электрическим током с плотностью $\uje$ [\si{\ampere\per\square\meter}]; а второе --- электрическое поле, созданное электрическими зарядами самих вакансий. Потенциал $\uPhi$ определяется уравнением Пуассона
\begin{equation}
\label{eq:Poisson}
\Delta\uPhi = - \frac{\uq}{\uepsepsZ} (\uC - \uC_0),
\end{equation}
где $\uC_0$ --- концентрация вакансий, соответствующая электрической нейтральности всего мемристора в целом, а $\uepsepsZ$ --- диэлектрическая проницаемость материала мемристора [\si{\farad\per\meter}]. Поскольку к мемристору подключены хорошо проводящие контакты (как схематически показано на вставке к нижней части Рис.~\ref{fig:profiles}), потенциал $\uPhi|_{\uX=0,\uL}=0$. До сих пор, в рамках данной модели мемристора, потенциал $\uPhi$ не учитывался и его учёт является главной особенностью данной работы.

Полное электрическое сопротивление мемристора в модели~\cite{boylo2020} рассчитывается феноменологически, учитывая (первый в разложении по степеням) линейный член в зависимости удельного сопротивления (объёмного и поверхностного на контактах) от локальной концентрации вакансий. В этом порядке, перераспределение вакансий в замкнутом мемристоре (с сохранением их общего числа) не приводит к изменению полного сопротивления его объёма (хотя и оказывает влияние на кинетику переключения сопротивления~\cite{boylo2025}), а в полном сопротивлении остаётся лишь поверхностный член $R=R_0+(R_1+R_2)\sigma$, где~\cite{boylo2025}
\begin{equation}
 \label{eq:resistance}
 \sigma=\frac{\uC|_{\uX=0}}{\uCmax} \cos^2\theta  + \frac{\uC|_{\uX=L}}{\uCmax} \sin^2\theta,
\end{equation}
где $R_0$, $R_1$, $R_2$ --- параметры материала мемристора и контактов [\si{\ohm}], а угол $0<\theta=\arctan \sqrt{R_2/R_1} < \pi/2$ характеризует асимметрию интерфейсов. Для того, чтобы сопротивление закрытого мемристора во ``включенном'' и ``выключенном'' состоянии отличалось, его интерфейсы (материалы и структура контактов) должны быть разными.

\section{Стационарные состояния мемристора}
К сожалению, точно решить полное кинетическое уравнение, вытекающее из~\eqref{eq:vacCurrent}, \eqref{eq:Poisson} пока не представляется возможным. Это уравнение подобно уравнению Бюргерса, но содержит нелокальный интегральный член, выражающий производную электростатического потенциала $\uphi$ через функцию Грина уравнения Пуассона. Подстановкой Коула-Хопфа это уравнение не линеаризуется.

Однако, стационарные состояния мемристора также представляют значительный интерес, поскольку именно они и определяют его сопротивления во ``включенном'' и ``выключенном'' состояниях. В закрытом мемристоре эти состояния соответствуют полному отсутствию тока вакансий $\uvJ=0$, что следует из уравнения непрерывности с учётом граничных условий. В рассматриваемом одномерном случае это сводится к следующей безразмерной системе уравнений
\begin{subequations}
\label{eq:stationary}
\begin{align}
 &-\ucx + \uc(1-\uc) \left(\up - \left(\alpha/2\right)\uphix\right) = 0, \label{eq:jzero}\\
 &\uphixx = - (c - r), \label{eq:phiPoisson}
\end{align}
\end{subequations}
где $\uc=\uC/\uCmax$ и $\uphi=\uPhi \uepsepsZ/(\uq\uCmax\uL^2)$ являются функциями безразмерной координаты $\ux=\uX/\uL$, безразмерный параметр $\up=2\uq\urho\uje\uL/(\ukBT)$ характеризует силу проходящей через мемристор плотности тока $\uje$, а безразмерный параметр $\ualpha = 4\uq^2\uL^2\uCmax/(\uepsepsZ\ukBT)>0$  характеризует силу взаимодействия зарядов вакансий в мемристоре. Коэффициент заполнения
\begin{equation}
 \label{eq:r}
 \ur=\int_0^1 c(x)\ud x
\end{equation}
является в закрытом мемристоре интегралом движения. При $\ualpha=0$ задача сводится к нахождению стационарного решения уравнения Бюргерса, которое хорошо известно
\begin{equation}
\uc|_{\ualpha=0}=\ucZ=\frac{e^{\up\ux}(e^{\up\ur}-1)}{e^{\up}-e^{\up\ur}-e^{\up\ux}+e^{\up(\ur+\ux)}}.
\end{equation}

Для нахождения решения при $\ualpha\neq0$ введём новую функцию (логит)
\begin{equation}
 \ul = \log\frac{\uc}{1-\uc},
\end{equation}
тогда уравнение для потока~\eqref{eq:jzero} линеаризуется и легко интегрируется ($\ul=p x - \ualpha\uphi/2 +\ulZ$), а уравнение~\eqref{eq:phiPoisson} превращается в
\begin{equation}
\ulxx = \frac{\ualpha}{2}\left(
\frac{e^\ul}{1+e^\ul} - \ur
\right).
\end{equation}
Это уравнение, хоть и нелинейное, но автономное, его аналитическое решение можно выразить в виде квадратуры
\begin{equation}
\label{eq:sol}
\int_{\ulZ}^\ul \frac{1}{\uAconst + \ualpha (\log (1+e^t) - \ur t )} \ud t = x,
\end{equation}
куда входят две постоянные интегрирования $\ulZ=\ul|_{x=0}$ и $\uAconst$. Вторую удобно определить в виде
\begin{equation}
 \uAconst = p^2\uupsilon^2 + \ualpha
 \left[\left(1-\ur\right)\log\left(1-\ur\right) + \ur \log \ur\right]
\end{equation}
через другую постоянную $\uupsilon$, которая имеет простой физический смысл и определена как
\begin{equation}
 \uupsilon = \frac{\ucx|_{\uc=\ur}}{\ucZx|_{\ucZ=\ur}}
\end{equation}
производная профиля распределения вакансий в точке $\uc=\ur$, нормированная на аналогичную производную, но в отсутствии электростатического взаимодействия ($\ualpha=0$). Вообще говоря, координаты $\ux$ этих двух точек с $\uc=r$ отличаются, как хорошо видно на Рис.~\ref{fig:profiles}. Можно показать, что всегда $0<\uupsilon<1$, что делает этот параметр особенно удобным. Полученное общее решение зависит от двух постоянных интегрирования $\ulZ$ и $\uupsilon$. Граничное условие для потенциала $\uphi(0)=0$ выполняется при этом автоматически, а из условия $\uphi(1)=0$ следует, что $\ul(1)-\ul(0)=p$. Второе уравнение для постоянных интегрирования получается из условия сохранения полного числа вакансий~\eqref{eq:r}. Решение для $\ulZ$ получается в явном виде
\begin{equation}
 \label{eq:ulZ}
 \ulZ = \log\frac{e^{\up\ur}-1}{e^\up-e^{\up\ur}},
\end{equation}
а уравнение для $\uupsilon$, которое получается из $\eqref{eq:sol}$ подстановкой туда $\ux=1$ и~\eqref{eq:ulZ}, удаётся решить только численно. Примеры получающихся профилей распределения вакансий показаны на Рис.~\ref{fig:profiles}, они удовлетворяют как уравнениям~\eqref{eq:stationary}, так и всем граничным условиям при заданном полном количестве вакансий в мемристоре. Можно получить и приближённое (но достаточно точное для оценок) явное выражение для $\uupsilon$, приведенное в Приложении~\ref{sec:upsapprox}.

\begin{figure}
\begin{center}
\includegraphics[width=1.0\columnwidth]{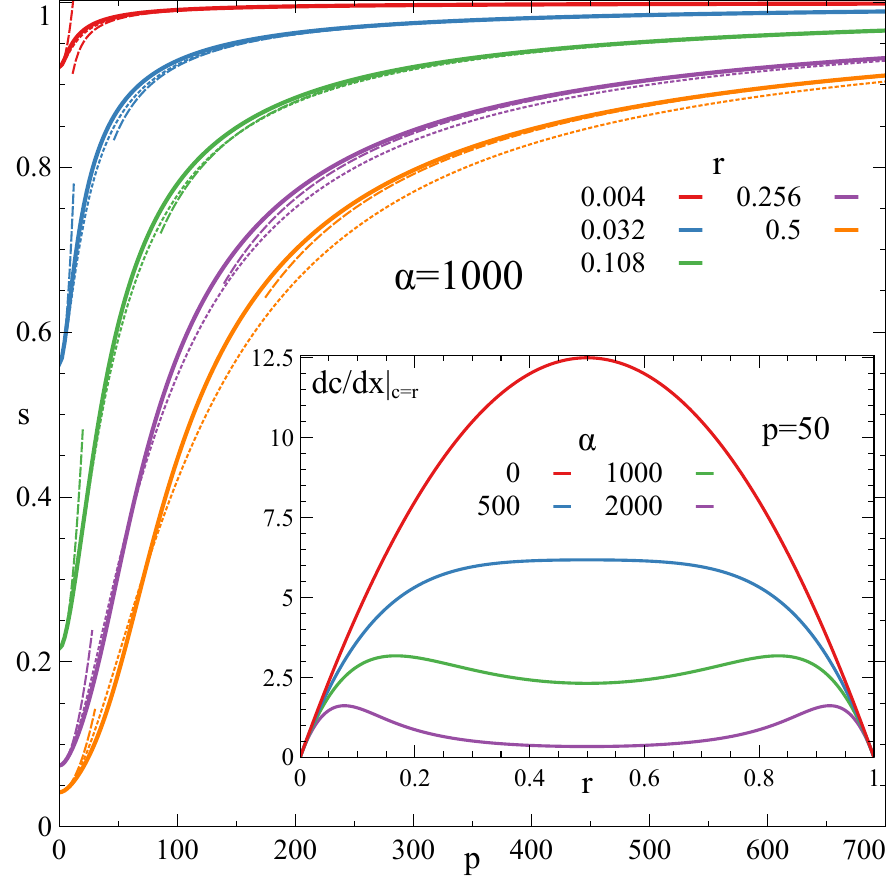}
\vspace{-0.8cm}
\end{center}
\caption{\label{fig:scale} Коэффициент масштабирования производной $s$ как функция приведенной силы тока $p$ в мемристорах с различным заполнением вакансиями при $\alpha=1000$. Кривые для заполнений $r$ и $1-r$ совпадают. На вставке показана зависимость производной профиля распределения вакансий $\ud c/\ud x$ в точке $c=r$ как функция коэффициента заполнения мемристора $r$ при $p=50$.}
\end{figure}

Зависимость коэффициента масштабирования производной $\uupsilon$ от силы тока $\up$ и от величины заполнения мемристора вакансиями при некоторых значениях величины электростатического взаимодействия $\ualpha$ показаны на Рис.~\ref{fig:scale}.
Эти зависимости симметричны относительно половинного заполнения (заполнение $\ur$ и $1-\ur$ приводят к одинаковым значениям $\uupsilon$) . По определению, чисто бюргерсовскому мемристору с $\ualpha=0$ соответствует $\uupsilon=1$. К этому же предельному значению стремится $\uupsilon$ при увеличении силы тока $\up$, эффект которого в конце концов превосходит эффект электростатического взаимодействия.

В целом, наличие электростатического взаимодействия приводит к тому, что в стационарном состоянии мемристора, взамен двух областей --- обогащенной и обеднённой вакансиями, формируется три -- обогащённая, обеднённая и промежуточная (тем более протяжённая, чем выше $\ualpha$) электрически нейтральная область. Похоже (хоть авторами это не выделялось) такая переходная область наблюдается в мемристорах на кислородных вакансиях в имплантированном ионами аргона рутиле~\cite{SVNK2023}.

Однако, несмотря на то, что распределение вакансий сильно меняется под влиянием электростатического взаимодействия между ними, на концентрацию вакансий на концах мемристора ($\uc(0)$ и $\uc(1)$) это взаимодействие не влияет, что следует~\eqref{eq:ulZ} из независимости $\ulZ$ от $\ualpha$. Это значит, что и предельные сопротивления (во ``включенном'' и ``выключенном'' состоянии) такого мемристора в линейном приближении~\eqref{eq:resistance}, от силы электростатического взаимодействия не зависят.

\section{Выводы}
Рассмотрена модель нелинейного мемристора на мобильных вакансиях в материале, где концентрация вакансий принимает значения в конечном фиксированном диапазоне. Учтены электростатическое взаимодействие между вакансиями. Получены аналитические выражения для стационарных (при некоторой величине проходящего через мемристор тока) распределений вакансий в мемристоре. Показано, что электростатическое взаимодействие не оказывает влияния на электрическое сопротивления мемристора во ``включенном'' и ``выключенном'' состояниях, по крайней мере в лидирующем линейном порядке по концентрации вакансий в объёме и на контактах.

\appendix
\renewcommand{\appendixname}{Приложение}
\renewcommand{\thesection}{\Asbuk{section}}
\section{Приближённое выражение для $\uupsilon$}
\label{sec:upsapprox}

Для $\uupsilon$ можно получить следующие асимптотические выражения:
\begin{align}
 \label{eq:upssmall}
 \upsilon (p\ll1) = & \uupsilon(0) + \frac{\uupsilon''(0)}{2!} p^2 + O(p^3), \\
 \upsilon(0) = & \frac{h}{\sinh h}, \\
 \upsilon''(0) = & \frac{h \left(8 h^2-\ualpha \right) (6 h \coth h-\cosh 2 h -5)}{288 \ualpha \sinh^3 h}, \\
 \label{eq:upslarge}
 \upsilon (p\gg1) = & 1 - \frac{2h^2}{p} + O(p^{-2}),
\end{align}
где $h=\sqrt{\ualpha \ur (1-\ur)}/(2\sqrt{2})$.

На их основе можно построить рациональную аппроксимацию
\begin{align}
 \label{eq:upspade}
 \uupsilon(p) \approx & \frac{h(576+a \up (1 + (\ur-1) \ur)))}{\sinh h (576+b \up (1 + (\ur-1) \ur)))}, \\
 a/b =& \frac{\sinh h \left(2 h^3-h \up+\up \sinh h\right)}{h \left(\left(2 h^2+\up\right) \sinh h-h \up\right)}, \\
 a-b =& 2 p-\frac{6 \up (h \coth h-1)}{\sinh^2 h},
\end{align}
которая хоть и не точно, но всё же достаточно хорошо приближает значения $\uupsilon$ во всей области параметров задачи. Асимптотики~\eqref{eq:upssmall}, \eqref{eq:upslarge} и приближённое выражение~\eqref{eq:upspade} показаны на Рис.~\ref{fig:scale} штриховыми линиями и точками.

%